\documentclass[12pt]{article}
\usepackage{amssymb}
\usepackage{amsmath}
\usepackage{latexsym}
\usepackage{array}
\usepackage{graphicx}
\usepackage{bm}
\usepackage{exscale}

\usepackage{theorem}
\theoremstyle{break}
\newtheorem{Theorem}{Theorem}%[section]
\newtheorem{Proposition}{Proposition}%[section]

\newtheorem{Corollary}{Corollary}
\newtheorem{Lemma}{Lemma}
\newtheorem{Definition}{Definition}%{section}
\def\qed{\hfill\hbox{$\Box$}\vspace{10pt}\break}

\begin{document}
\title{Singularities of the discrete KdV equation and the Laurent property}
\author{Masataka Kanki$^1$, Jun Mada$^2$, and Tetsuji Tokihiro$^1$\\
\small $^1$ Graduate School of Mathematical Sciences,\\
\small University of Tokyo, 3-8-1 Komaba, Tokyo 153-8914, Japan\\
\small $^2$ College of Industrial Technology,\\
\small Nihon University, 2-11-1 Shin-ei, Narashino, Chiba 275-8576, Japan}
\date{}
\maketitle

\begin{abstract}
We study the distribution of singularities for partial difference equations, in particular, 
the bilinear and nonlinear form of the discrete version of the Korteweg-de Vries (dKdV) equation.
Using the Laurent property, and the irreducibility, and co-primeness of the terms of the bilinear dKdV equation,
we clarify the relationship of these properties with the appearance of zeros in the time evolution.
The results are applied to the nonlinear dKdV equation and we formulate the famous integrability criterion (singularity confinement test) for nonlinear partial difference equations with respect to the co-primeness of the terms.

{\tt PACS2010: 45.05.+x, 02.30.Ik, 02.40.Xx, 02.30.Jr}

{\tt MSC2010: 37K10, 35A20, 35Q53, 47A07}

{\tt Keywords: singularity confinement, discrete KdV equation, coprime}
\end{abstract}

%%%%%%%%%%%%%%%%%%%%%%%%%%%%%%%%%%%%%%%%%%%%%%%%%%
\section{Introduction}
\label{sec1}
In the case of continuous equations, the Painlev\'{e} property is a useful criterion for integrability \cite{Conte}. Motivated by this property,
Grammaticos, Ramani and Papageorgiou introduced the
`singularity confinement' test \cite{SC}, which is quite useful as an integrability detector for ordinary difference equations.
They have discovered that there is a difference between the singularity structures of integrable and non-integrable discrete systems.
In the case of integrable systems, because of a fine cancellation of terms,
even if we start from infinity, we are lead to finite values after several steps. On the other hand, in non-integrable equations, singularities radiate out from a singular point.
Singularity confinement has been successfully utilized to identify discrete Painlev\'{e} equations.
Motivated by the results of the singularity confinement test, Sakai has completely classified the discrete Painlev\'{e} equations as bi-rational maps on rational surfaces obtained by blowing up two dimensional projective spaces \cite{Sakai}.  
Recently, it was discovered that the notion of singularity confinement can be appropriately reformulated even for equations over the field of $p$-adic numbers and the compatibility with their reduction modulo a prime number, and applied to the equations over finite fields \cite{Kanki,Kanki2}.

The singularity confinement approach was also taken for partial difference equations in their bilinear forms by Ramani, Grammaticos and Satsuma \cite{RGS}. 
In their approach, singularity is defined as the zeros of the dependent variables and its confinement means that zeros will not propagate in general situations. 
The aim in the present paper is to formulate the confinement of singularities for general nonlinear partial difference equations in a rigorous manner.
Since it is not practical to go through all the configurations of singular points one by one, it is desirable to redefine the confinement of singularities in terms of the algebraic or analytic relations between adjacent terms.
For this purpose, we utilize the so-called \textit{Laurent property}.
An equation has the Laurent property if, for a given initial condition, the solution of the equation is expressed by Laurent polynomials in these initial data. It has already been proved
by Fomin and Zelevinsky that bilinear forms of the Hirota-Miwa equation and the discrete KdV (dKdV) equation have the Laurent property using the concept of cluster algebras \cite{FZ,FZ2}. Recently, T. Mase further investigated this area and obtained general criteria for the existence of Laurent properties for various boundary conditions and reductions \cite{Mase}.
A.N.W. Hone investigated the relationship between the singularity confinement and the Laurent property. He presented many non-integrable mappings that have the Laurent property and also have confined singularities \cite{Hone}.
From these previous results, we have learned that, although Laurent property is a prospective integrability criterion, it does not perfectly identify integrable mappings. Therefore in this article we also focus on other properties: the irreducibility and co-primeness of the distinct terms.
Our assertion is that this co-primeness is the mathematical re-interpretation of confinement of singularities for partial difference equations. 
To be specific, we consider the dKdV equation in both bilinear and nonlinear forms. 
It is to be expected that a similar discussion should be possible for general partial difference equations.
To make this paper self-contained and to fix notations, we first prove the Laurent property of the bilinear form of the dKdV equation in an elementary way. 
The irreducibility of the terms and the co-primeness of the distinct terms is proved in the course of this proof.
Then we investigate the nonlinear dKdV equation by utilizing these results. 
The main theorem in this paper is that the distinct terms of the nonlinear dKdV equation do not have common factors other than monomials, if they are separated by more than one cell.
We conclude that this theorem describes how the singularities are confined for the discrete KdV equation.

\section{Singularity of dKdV equation}
The bilinear form of the dKdV equation is given by
\begin{equation}
(1+\delta)a_{m-1}^{n-1} a_m^{n+1} = a_{m-1}^n a_m^n + \delta a_{m-1}^{n+1} a_m^{n-1}, \label{bilin}
\end{equation}
where $m$ and $n$ are integer independent variables.
Initial conditions are given by designating the values for the following set $I_a$, consisting of two rows and one column:
\[
I_a=\{a_m^0,\ a_m^1,\ a_0^n | m\ge 0, n\ge 0\}.
\]
The nonlinear form of the discrete KdV equation is given by
\begin{equation}
\frac{1}{w_{m+1}^{n+1}}-\frac{1}{w_m^n}+\frac{\delta}{1+\delta}(w_m^{n+1}-w_{m+1}^n)=0, \label{nonlin}
\end{equation}
where the initial conditions are given by the set $I_w$:
\[
I_w=\{w_m^0,\ w_0^n | m\ge 0, n\ge 0\}.
\]
\begin{figure}
\centering
\includegraphics[width=12cm, bb=100 550 500 750]{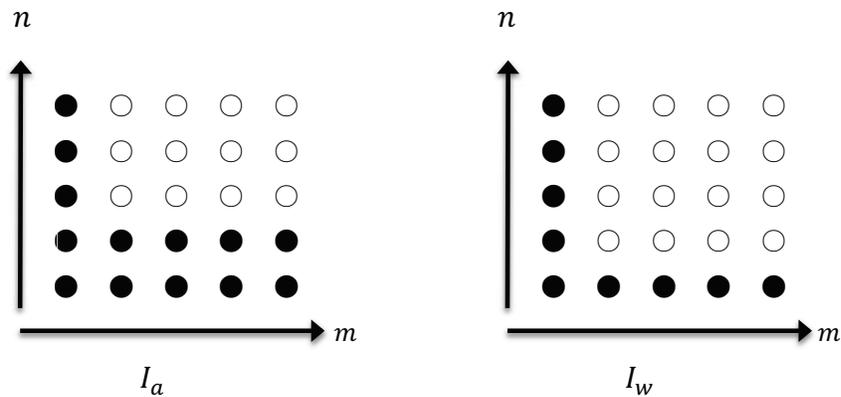}
\caption{Initial conditions $I_a$ and $I_w$ of the bilinear dKdV \eqref{bilin} and nonlinear dKdV \eqref{nonlin}. Filled circles `$\bullet$' indicate the initial conditions, while the values at open circles `$\circ$' are determined by the evolution equation \eqref{bilin} or \eqref{nonlin} from these initial conditions.}
\label{figure1}
\end{figure}
The initial conditions $I_a$ and $I_w$ are shown on the $m$-$n$ plane in the figure \ref{figure1}.
These two equations are connected by the relation
\begin{equation}
w_m^n=\frac{a_{m-1}^{n+1} a_m^n}{a_m^{n+1} a_{m-1}^n}. \label{relation}
\end{equation}
We consider the evolution of these equations on the field $K=\mathbb{Q}(\delta)$. The ring of Laurent polynomials with respect to the elements of a set $I$ with coefficients in the field $K$ is denoted by
\[
K[I]:=K[f^{\pm 1} | f\in I],
\]
and we denote the set of all monomials in $K[I]$ as $M[I]$.
\begin{Definition}
Two Laurent polynomials $f,g\in K[I]$ are co-prime if
every common factor is at most a monomial:
\[
f=h f',\ g=h g'\ \Rightarrow \ h\in M[I].
\]
If $f\in K[I]$ is co-prime with every element $g\in K[I]$ such that $g\not \in M[I]f:=\{mf|\ m\in M[I]\}$, then
$f$ is said to be an irreducible Laurent polynomial.
We denote the set of all irreducible Laurent polynomial in $K[I]$ as $K_0[I]$.
\end{Definition}

First we shall prove Proposition \ref{laurentbilin}, which establishes the Laurent properties and co-primeness of equation \eqref{bilin}.
The Laurent property of equation \eqref{bilin} was first proved by Fomin and Zelevinsky using the so-called `caterpillar lemma' associated with cluster algebras \cite{FZ2}.  
Here we give an elementary proof of the Laurent property for dKdV equation without using the notion of a cluster algebra
and show the co-primeness in the course of the proof.
In the original work by Fomin and Zelevinsky, the coefficients of the Laurent polynomials are taken in the ring of integers.
In our paper, since we are dealing with functions with coefficients in $K=\mathbb{Q}(\delta)$, we do not have to discuss integerness of coefficients. Instead, the Laurent property, the irreducibility and co-primeness are studied over the field $K$.
Let us mention one fact about the integerness of equation \eqref{bilin} without going into the details. From proposition \ref{laurentbilin}, we can prove that $a_m^n\in\mathbb{Z}$ for all $m\ge 1$ and $n\ge 1$, if we take $x_m=1, y_n=1$ for all $m, n\ge 1$.
\begin{Lemma}\label{kiyakulemma}
Let us assume that $g_1,g_2\in K[I]\setminus\{0\}$ are co-prime,
and that the variable $x$ does not appear in $g_1$ nor in $g_2$.
Then $f=x g_1+g_2$ is irreducible $(f\in K_0[I])$.
Moreover, for every $g\in K[I]$ that does not depend on $x$, $f$ is co-prime with $g$.
\end{Lemma}
\textbf{Proof}\;\;
Let us assume that $f$ is factored as $f= h_1 h_2$ $(h_1, h_2\in K[I])$.
Since $f$ is a polynomial of degree one in terms of $x$,
we assume that $h_1$ is of degree one in $x$ and that $h_2$ is of degree zero. Then $h_2$ divides both of $g_1$ and $g_2$ in $K[I]$.
Since $g_1$ and $g_2$ are co-prime, $h_2\in M[I]$.
Therefore $f\in K_0[I]$.
Next let us take a $g\in K[I]$ that does not contain $x$.
If $f$ and $g$ are not co-prime, there exist
$h\in K[I]\setminus M[I]$ and $f',g'\in K[I]$ such that $f=hf'$ and $g=hg'$. Since $f$ is irreducible ($f\in K_0[I]$), we have that $f'\in M[I]$. Therefore $g=\frac{g'}{f'} f$ where $g'/f'$ is a Laurent polynomial.
Since $f$ is of degree one in terms of $x$,  and the constant term of $f$ (i.e. $g_2$) is not zero, $g$ must be dependent on $x$, which leads to a contradiction.
\qed

From here on, let us take the initial conditions for the bilinear dKdV equation as
\begin{equation}
a_m^0=1\ (m\ge 0),\ a_m^1=x_m\ (m\ge 1),\ a_0^n=y_n\ (n\ge 1) \label{initcondition}.
\end{equation}
These conditions do not seem to be general ones, however, variables $x_m,y_n$ are enough to ensure the general initial conditions for the corresponding nonlinear dKdV equation \eqref{nonlin} and its initial conditions $I_w$.
Hence we have $I_a=\left\{\{x_m\}_{m=1}^\infty,\,\{y_n\}_{n=1}^\infty  \right\}$.

\begin{figure}
\centering
\includegraphics[width=9cm, bb=100 600 400 750]{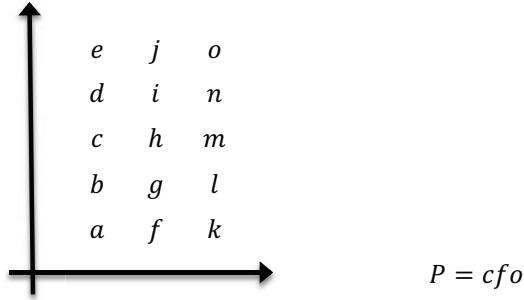}
\caption{Configuration of the renamed variables $a:=a_m^n$ through $o:=a_{m+4}^{n+4}$ of the bilinear dKdV equation \eqref{bilin} in the $m$-$n$ plane.
The function $P$ in the lemma \ref{polynomiallemma} is equal to $cfo$.}
\label{figure2}
\end{figure}

\begin{Lemma} \label{zerolemma}
The variable $a_m^n$ is a rational function of $x_i$ ($1 \le i \le m$) and $y_j$ ($1 \le j \le n$).
For every $m,n\in \mathbb{Z}$, $a_m^n$ is not identically zero. 
\end{Lemma}
\textbf{Proof}\;\;
The former statement is inductively proved from equation \eqref{bilin}. 
If we substitute $x_m=1,\ y_n=1\ (m,n\ge 1)$, then we obtain that $a_m^n=1$ for all $m,n\in\mathbb{Z}$,
since $a_m^n=1(\forall m \forall n)$ is a trivial solution for the equation \eqref{bilin}.
Since there is at least one set of initial conditions where $a_m^n=1\neq 0$, we conclude that $a_m^n$ cannot be identically equal to zero as functions of $x_m$ and $y_n$.\qed
\begin{Lemma}\label{3row}
Let us fix $m\ge 2$.
If we suppose $a_k^2\in K[I_a]\ (\forall k\le m)$ and $a_{k}^3\in K[I_a]\ (\forall k\le m-1)$, then
we have $a_m^3\in K_0[I_a]$.
\end{Lemma}
\textbf{Proof}\;\;
From equation \eqref{bilin}, we have
\[
(1+\delta) a_m^3 x_{m-1}= \delta x_m a_{m-1}^3 + a_m^2 a_{m-1}^2.
\]
Therefore
\[
\frac{a_m^3}{x_m}=\frac{\delta}{1+\delta}\frac{a_{m-1}^3}{x_{m-1}}+\frac{1}{1+\delta}\frac{a_m^2 a_{m-1}^2}{x_m x_{m-1}}.
\]
By induction, we obtain
\begin{equation*}
\frac{a_m^3}{x_m}=\left(\frac{\delta}{1+\delta}\right)^m \frac{y_3}{y_1}+A,
\end{equation*}
where
\[
A=\frac{\delta^{m-1}}{(1+\delta)^m}\frac{a_1^2 y_2}{x_1 y_1}+\frac{\delta^{m-2}}{(1+\delta)^{m-1}}\frac{a_2^2 a_1^2}{x_2 x_1}+\cdots +\frac{1}{1+\delta} \frac{a_m^2 a_{m-1}^2}{x_m x_{m-1}}.
\]
The terms  in $A$ do not contain $y_3$.
We also observe that $A$ is not identically zero, since if we take $x_k=y_k=1 (k\in\mathbb{Z})$ in $A$, then all the terms in $A$ are positive rational numbers.
Therefore from lemma \ref{kiyakulemma}, we have $a_m^3\in K_0[I_a]$.
\qed
\begin{Lemma}\label{allrow}
Let us fix $(m,n)\in\mathbb{Z}_{>0}^2$.
Let us assume that $a_k^l\in K[I_a]$ for $\forall k\le m,\ \forall l \le n$, and that $a_k^l\in K_0[I_a]$ for $\forall k\le m, \forall l\le n$ with $(k,l)\neq (m,n)$. Then we have $a_m^n\in K_0[I_a]$.
\end{Lemma}
\textbf{Proof}\;\;
We only have to prove this lemma for $n\ge 4$, since the case of $n=3$ is proved in lemma \ref{3row} and the cases of $n=0,1,2$ are trivial.
In a similar manner to the proof of lemma \ref{3row}, we obtain
\begin{equation}
a_m^n = \left(\frac{\delta}{1+\delta}\right)^m \frac{y_n}{y_{n-2}}a_m^{n-2}+a_m^{n-2} g, \label{lem4eq}
\end{equation}
\begin{equation*}
g=\frac{\delta^{m-1}}{(1+\delta)^m}\frac{a_1^{n-1} y_{n-1}}{a_1^{n-2} y_{n-2}}+\frac{\delta^{m-2}}{(1+\delta)^{m-1}}\frac{a_2^{n-1} a_1^{n-1}}{a_2^{n-2} a_1^{n-2}}+\cdots +\frac{1}{1+\delta} \frac{a_m^{n-1} a_{m-1}^{n-1}}{a_m^{n-2} a_{m-1}^{n-2}}.
\end{equation*}
If we put $p_k=a_k^{n-2}$, $q_k=a_k^{n-1}$, $(k=1,2,\cdots ,m)$ for simplicity, then there exists a Laurent polynomial $f\in K[I_a]$ such that
\begin{equation*}
p_m g=\frac{D\cdot f\cdot p_m+ y_{n-2}p_1 \cdots p_{m-2} q_{m-1} q_m}{(1+\delta) y_{n-2}p_1 p_2\cdots p_{m-1}},
\end{equation*}
where $D\in K$.
Let us note that $g$ does not contain the variable $y_n$.
From the assumption that $a_m^n\in K[I_a]$ and equation \eqref{lem4eq}, we have $p_m g\in K[I_a]$.
We consider $a_m^n$ as a polynomial of $y_n$ of degree one, and show that coefficient $p_m/y_{n-2}$ of $y_n$, and the constant term $p_m g$ are co-prime.
Since $y_{n-2} p_1 \cdots p_{m-2} q_{m-1} q_m$ is co-prime with $p_m$ by an assumption that $p_m\in K_0[I]$, the numerator of $p_m g$ is not divisible by $p_m$ in the ring of Laurent polynomials $K[I]$. Thus $p_m g$ and $p_m$ does not have common factors other than monomials. Therefore $p_m/y_{n-2}$ and $p_m g$ are co-prime. From lemma \ref{kiyakulemma} we have that $a_m^n$ is irreducible (i.e. $a_m^n\in K_0[I_a]$).
\qed
\begin{Lemma}\label{tagainiso}
If $a_m^n\in K_0[I_a]$ and $a_k^l\in K_0[I_a]$ for $(m,n)\neq (k,l)$, then $a_m^n$ and $a_k^l$ are co-prime.
\end{Lemma}
\textbf{Proof}\;\;
We assume that $l\le n$. First, if $l<n$, by a calculation similar to that in lemma \ref{allrow}, we observe that
$a_m^n$ has the form $a_m^n=y_n g_1 +g_2$, where $g_1, g_2\in K[I_a]\setminus \{0\}$. Since $a_k^l\, (l<n)$ does not contain $y_n$, $a_m^n$ and $a_k^l$ are co-prime from lemma \ref{kiyakulemma}.
Next we prove the case of $l=n$. We can assume that $k<m$.
We suppose that $n\ge 2$, as the cases of $n=0,1$ is trivial.
We will show that $a_m^n$ and $a_k^n$ are co-prime. Since $a_k^n$ does not contain $x_m$, while $a_m^n$ does, we conclude that $a_m^n /a_k^n\in M[I_a]$ and that $a_m^n /a_k^n$ has a positive order of $x_m$ as one of its factors. Therefore if we take $x_m=0$ then $a_m^n=0$.
However, since $a_m^2\neq 0$, $a_{m-1}^2\neq 0$ under the condition $x_m=0$, we have to conclude that $a_m^n\neq 0$ from the recurrence \eqref{bilin},
which leads to a contradiction.
\qed
\begin{Lemma}\label{polynomiallemma}
For every $(m,n)\in \mathbb{Z}_{>0}^2$, the quantity
\[
P:=a_{m+1}^n a_m^{n+2} a_{m+2}^{n+4}
\]
is a polynomial in $a_i^j$ where $m\le i\le m+2$, $n\le j\le n+4$ and $(i,j)\neq (m+2,n+4)$, i.e.
\[
P\in K[a_i^j|\ m\le i\le m+2,\ n\le j\le n+4,\ (i,j)\neq (m+2,n+4)].
\]
\end{Lemma}
\textbf{Proof}\;\;
For simplicity we rename the variables as
\begin{eqnarray*}
(a_m^n, a_m^{n+1}, \cdots, a_m^{n+4})&=&(a,b,c,d,e),\\
(a_{m+1}^n, a_{m+1}^{n+1}, \cdots, a_{m+1}^{n+4})&=&(f,g,h,i,j),\\
(a_{m+2}^n, a_{m+2}^{n+1}, \cdots, a_{m+2}^{n+4})&=&(k,l,m,n,o).
\end{eqnarray*}
See figure \ref{figure2}.
We show that $c\cdot f \cdot o$ can be written as a polynomial in $a$ through $n$, i.e.: without using $o$.
First we compute $cfjm$.
\begin{eqnarray*}
cfjm&=&(cj)(fm)=\frac{\delta eh +di}{1+\delta}\frac{\delta hk +gl}{1+\delta}\\
&=&\frac{1}{(1+\delta)^2}\left( dgil + h(\delta egl +\delta dik +\delta^2 ehk) \right)\\
&=&\frac{1}{(1+\delta)^2}(dg)(il)+\frac{h(\delta egl +\delta dik +\delta^2 ehk)}{(1+\delta)^2}\\
&=&\frac{1}{(1+\delta)^2}\frac{(1+\delta) bi -ch }{\delta}\frac{(1+\delta) gn -hm}{\delta}+\frac{h(\delta egl +\delta dik +\delta^2 ehk)}{(1+\delta)^2}\\
&=&\frac{1}{\delta^2} bgin + \frac{1}{(1+\delta)^2} h\cdot X,
\end{eqnarray*}
where
\[
X=-\frac{1+\delta}{\delta^2} cgn -\frac{1+\delta}{\delta^2} bim +\frac{1}{\delta^2} chm +\delta egl  +\delta dik +\delta^2 ehk.
\]
Here we have used equation \eqref{bilin} several times.
Furthermore, we obtain
\begin{eqnarray*}
cfoh&=&cf \left(\frac{in +\delta jm}{1+\delta}\right)=\frac{1}{1+\delta}(cfin +\delta\cdot cfjm)\\
&=&\frac{1}{1+\delta}\left( cfin + \frac{1}{\delta} bgin+\frac{\delta}{(1+\delta)^2} h\cdot X \right)\\
&=&\frac{bg +\delta cf}{\delta(1+\delta)}in +\frac{\delta}{(1+\delta)^3}h X=\left( \frac{1}{\delta}ain +\frac{\delta}{(1+\delta)^3} X \right) h.
\end{eqnarray*}
Since $h\neq 0$, we have shown that $a_{m+1}^n a_{m}^{n+2} a_{m+2}^{n+4} =cfo$ can be written as a polynomial in $a,b,c,\cdots, n$, i.e.: without using $o$. 
\qed
\begin{Proposition} \label{laurentbilin}
The bilinear dKdV equation \eqref{bilin} with initial condition \eqref{initcondition} satisfies the following:
\begin{itemize}
\item $a_m^n\in K_0[I_a]$,
\item $a_m^n$ and $a_{m'}^{n'}$ are co-prime if $(m,n)\neq (m',n')$.
\end{itemize}
\end{Proposition}
\textbf{Proof}\;\;
We prove proposition \ref{laurentbilin} by induction.
(i) First let us prove the case $m=1$. We know that $a_1^2=(x_1 y_1+\delta y_2)/(1+\delta)$ is irreducible and is co-prime with both $a_1^0=1$ and $a_1^1=x_1$.
Let us take a positive integer $k$ and suppose that $a_1^n\in K_0[I_a]$ for $n=1,2,\cdots, k$ and suppose that
$a_1^n$ and $a_1^{n'}$ are co-prime for every $n,n'\le k$.
The evolution equation \eqref{bilin} implies that
\[
a_1^{k+1}=\frac{y_k a_1^k+\delta y_{k+1} a_1^{k-1}}{(1+\delta) y_{k-1}},
\]
where $a_1^{k-1},a_1^{k}\in K_0[I_a]$.
Since neither $a_1^k$ nor $a_1^{k-1}$ contains $y_{k+1}$, by lemma \ref{kiyakulemma}, we conclude that $a_1^{k+1}$ is irreducible and that all $a_1^n (n=1,\cdots ,k+1)$ are co-prime. This settles the case $m=1$.

(ii) Next we move on to the case of $m\ge 2$.
Let us fix $l\ge 2$ and suppose that proposition \ref{laurentbilin} holds for every $1\le m\le l$ and every $n\ge 1$.
(ii-a) First we prove that $a_{l+1}^2, a_{l+1}^3\in K_0[I_a]$.
By equation \eqref{bilin},
\[
(1+\delta ) a_{l+1}^2= x_l x_{l+1} +\delta a_l^2.
\]
By the assumption that $a_l^2$ is irreducible and the fact that $a_l^2$ does not contain $x_{l+1}$, using lemma \ref{kiyakulemma}, we have $a_{l+1}^2\in K_0[I_a]$. Since
\[
a_{l+1}^3=\frac{\delta x_{l+1} a_l^3 + a_l^2 a_{l+1}^2}{(1+\delta ) x_l},
\]
we obtain $a_{l+1}^3\in K[I_a]$. Thus from lemma \ref{3row}, we conclude that $a_{l+1}^3\in K_0[I_a]$.
(ii-b) Next, let us prove by induction that $a_{l+1}^n\in K_0[I_a]$ for all $n\ge 4$. Let us fix $k\ge 3$. We suppose that $a_{l+1}^k\in K_0[I_a]$ and prove that $a_{l+1}^{k+1}\in K_0[I_a]$.
By equation \eqref{bilin},  we have
\[
(1+\delta ) a_{l+1}^{k+1} a_l^{k-1} = \delta a_l^{k+1} a_{l+1}^{k-1} + a_l^k a_{l+1}^k.
\]
Therefore $a_{l+1}^{k+1} a_l^{k-1}$ is a Laurent polynomial by the assumption.
From lemma \ref{polynomiallemma}, we show that there exists a polynomial $P$ (therefore of course $P\in K[I_a]$) such that
\[
a_l^{k-3} a_{l-1}^{k-1} (a_{l+1}^{k+1} a_l^{k-1}) = a_l^{k-1} P.
\]
From lemma \ref{tagainiso}, terms $a_l^{k-3}$, $a_{l-1}^{k-1}$ and $a_l^{k-1}$ are all co-prime.
Therefore both $a_l^{k-3}$ and $a_{l-1}^{k-1}$ are factors of $P$.
Thus the term
\[
a_{l+1}^{k+1}=\frac{P}{a_l^{k-3} a_{l-1}^{k-1}}
\]
is a Laurent polynomial.
Then from lemma \ref{allrow}, we find that the term $a_{l+1}^{k+1}$ is an irreducible Laurent polynomial.
Therefore we have proved by induction that $a_{l+1}^n\in K_0[I_a]$ for all $n\ge 1$.
The co-primeness of the terms follows directly from lemma \ref{tagainiso}.
\qed
Next we investigate the \textit{nonlinear} dKdV equation. To  enable us to easily apply the previous proposition to the nonlinear dKdV equation, we need a `simple' relationship between the nonlinear and bilinear equations, preferably one that can be expressed by only monomials.
For this purpose, in the bilinear dKdV equation, we limit ourselves to the case of
\[
y_n=1\ (n\ge 1),
\]
in addition to the initial condition \eqref{initcondition}.
The correspondence between the initial values $I_w$ of nonlinear dKdV equation and $I_a$ of bilinear dKdV equation is obtained as:
\begin{equation} \label{nonlinbilinrel}
x_1=\frac{1}{w_1^0},\ x_2=\frac{1}{w_1^0 w_2^0},\cdots,\ x_m=\frac{1}{\prod_{k=1}^m w_k^0}.
\end{equation}
% Note that suppose we \textit{do not} have monomial relation like \eqref{nonlinbilinrel}, then the irreducibility and co-primeness of
% the bilinear dKdV equation \textit{do not} indicate those properties of the nonlinear dKdV equation \eqref{nonlin}.
Let $L_m$ be a ring of Laurent polynomials of $m$ variables ($x_1$ through $x_m$):
\[
L_m:=K[x_1^{\pm 1}, \cdots, x_m^{\pm 1}].
\]
\begin{Lemma}\label{xmlemma}
\begin{itemize}
\item $a_m^n$ is a one-dimensional polynomial in $x_m$,
\[
a_m^n=p_m(n) x_m +q_m(n),
\]
where $p_m(n), q_m(n)\in L_{m-1}$.

\item
In particular, if we take $m=1$,
we have $p_1(n)+q_1(n)=1$. We also have
$p_1(n)\neq p_1(n')$ and $q_1(n)\neq q_1(n')$ for $n\neq n'$.
\end{itemize}
\end{Lemma}
\textbf{Proof}\;\;
The first part is immediate, because the Laurent property of the coefficients $p_m(n)$ and $q_m(n)$ was already proven in proposition \ref{laurentbilin}, and the form of $a_m^n$ is easily obtained by induction.
To prove the second part we take $m=1$ and give the recurrence relations for $p_1(n)$ and $q_1(n)$. For simplicity we omit the subscripts and write $p(n):=p_1(n),\ q(n):=q_1(n)$. The recurrence relations are
\begin{eqnarray*}
p(n+2)&=&\frac{1}{1+\delta}(p(n+1)+\delta p(n)),\\
q(n+2)&=&\frac{1}{1+\delta}(q(n+1)+\delta q(n)),
\end{eqnarray*}
where the initial conditions are $p(0)=0, p(1)=1, q(0)=1, q(1)=0$.
By solving these relations we obtain
\[
p(n)=\frac{1+\delta}{1+2\delta}\left\{1-\left(\frac{-\delta}{1+\delta}\right)^n\right\},\; q(n)=\frac{\delta}{1+2\delta} \left\{1-\left(\frac{-\delta}{1+\delta}\right)^{n-1}\right\}.
\]
It implies that $p(n)+q(n)=1$. We also observe that $p(n)\neq p(n')$ and $q(n)\neq q(n')$ for $n\neq n'$ as elements of $K$.
\qed

%%%%%%%%%%%%%%%%%%%%%%%%%%%%%%%%%%%%%%%%%%%%%%%%%%
\begin{Proposition} \label{prop2}
The bilinear dKdV equation \eqref{bilin} with the restricted initial condition
\[
a_m^0=1,\ a_0^n=1,\ a_k^1=x_k\ (m,n\ge 0,\ k\ge 1),
\]
satisfies the following:
\begin{itemize}
\item $a_m^n\in L_m$,
\item $a_m^n$ is irreducible as an element of $L_m$,
\item $a_m^n$ and $a_{m'}^{n'}$ are co-prime as elements of $L_m$ if $(m,n)\neq (m',n')$.
\end{itemize}
\end{Proposition}
\textbf{Proof}\;\; 
The first assertion that every term is a Laurent polynomial is already proved in proposition \ref{laurentbilin}.
However, the irreducibility and co-primeness are not necessarily conserved under the substitutions $y_1=1, y_2=1, \cdots$. Therefore we need a different proof
for these two properties.
First we prove the case of $m=1$.
If we suppose that $a_1^k$ and $a_1^n$ are not co-prime for some $k\neq n$, then there exists $f\in K$ such that $a_1^k=f a_1^n$.
Thus we have $p(k)x_1+q(k)=(p(n)x_1+q(n))f$.
From the latter half of lemma \ref{xmlemma},
\[
\frac{q(n)}{q(k)}=\frac{p(n-1)}{p(k-1)}\neq \frac{p(n)}{p(k)}
\]
for $n\neq k$. Therefore we conclude that $n=k$ and $f=1$.
This settles the case $m=1$.

Next we prove the case of $m\ge 2$ by induction.
Let us suppose that proposition \ref{prop2} holds for $m\le k$ and let us prove that $a_{k+1}^n$ is irreducible for every $n\ge 0$.
Since $a_{k+1}^0=1, a_{k+1}^1=x_{k+1}$, the case of $n=0,1$ is trivial.
Let us suppose that $a_{k+1}^n$ is \textit{reducible} for $n\le l$ and prove the case of $n=l+1$.
If $a_{k+1}^{l+1}$ is \textit{reducible}, two terms $p_{k+1} (l+1)$ and $q_{k+1} (l+1)$ are not co-prime. We pick one common factor $f\in L_k$ that
is irreducible and is not a monomial.
Let us define two matrices $P_m^n$ and $A_m^n$ as
\[
P_m^n=\begin{pmatrix}
p_m(n) & q_m(n) \\
p_m(n-1) & q_m(n-1)
\end{pmatrix}
,\;\;
A_m^n=\begin{pmatrix}
\alpha_m (n) & \beta_m (n) \\
1 & 0
\end{pmatrix}
\]
where
\[
\alpha_m (n)=\frac{a_{m-1}^{n-1}}{(1+\delta) a_{m-1}^{n-2}},\ \beta_m (n)=\frac{\delta a_{m-1}^n}{(1+\delta) a_{m-1}^{n-2}}.
\]
Then we have $P_m^n=A_m^n P_m^{n-1}$ from equation \eqref{bilin}.
Therefore,
\begin{equation}
P_m^n=A_m^n A_m^{n-1} \cdots A_m^2. \label{matrixP}
\end{equation}
Since $f$ is a common factor of $p_{k+1}(l+1)$ and $q_{k+1}(l+1)$, the determinant $\det P_{k+1}^{l+1}\equiv 0 \mod f$.
From \eqref{matrixP},
\begin{eqnarray*}
0&\equiv& \det (A_{k+1}^{l+1} A_{k+1}^l\cdots A_{k+1}^2)\equiv (-1)^l \beta_{k+1}(l+1)\cdot \beta_{k+1}(l)\cdots \beta_{k+1}(2)\\
&\equiv& \left(\frac{-\delta}{1+\delta}\right)^l \frac{a_k^{l+1} a_k^l}{x_k} \mod f.
\end{eqnarray*}
This implies that $a_{k}^{l+1} a_k^l\equiv 0 \mod f$.
From \eqref{bilin},
\[
(1+\delta) a_k^{l-1} a_{k+1}^{l+1}=a_k^l a_{k+1}^l +\delta a_k^{l+1} a_{k+1}^{l-1}.
\]
From $a_{k+1}^{l+1}=p_{k+1}(l+1)x_{k+1} + q_{k+1}(l+1)\equiv 0\mod f$,
we observe
\begin{equation}
a_k^l a_{k+1}^l\equiv -\delta a_k^{l+1} a_{k+1}^{l-1}\mod f. \label{modf}
\end{equation}
From the assumption that both $a_{k+1}^{l-1}$ and $a_{k+1}^l$ are irreducible and from the fact that $f$ does not contain $x_{k+1}$, while $a_{k+1}^{l-1}$ and $a_{k+1}^l$ do, we have that
$f$ is not a factor of either $a_{k+1}^{l-1}$ or $a_{k+1}^l$.
(We have used lemma \ref{kiyakulemma} here.)
Therefore, from \eqref{modf}, we conclude that 
\[
a_k^{l+1}\equiv a_k^l\equiv 0\mod f.
\]
This contradicts the assumption that $a_k^l$ and $a_k^{l+1}$ are co-prime.
Thus we have proved that $a_{k+1}^{l+1}$ is irreducible.

Finally we prove that $a_{k+1}^{l+1}$ is co-prime with $a_m^n$ for $m\le k+1$, $n\le l+1$, $(m,n)\neq (k+1,l+1)$.
First if $m\le k$ then, since $a_m^n$ does not contain $x_{k+1}$, the term $a_m^n$ is co-prime with $a_{k+1}^{l+1}$.
Next, let us take $m=k+1$ and prove that $a_{k+1}^n\ (n\le l)$ and $a_{k+1}^{l+1}$ are co-prime. If we suppose otherwise, there exists $f\in M[I_a]$ such that $a_{k+1}^{l+1}=f a_{k+1}^n$, since irreducible Laurent polynomials can only have monomials as common factors.
If we take $x_1=x_2=\cdots=x_k=1$ in the variables, then $f|_{x_1=\cdots=x_k=1}=1$ from lemma \ref{xmlemma}. Therefore $a_{k+1}^{l+1}=a_{k+1}^n$ under the condition $x_1=x_2=\cdots=x_k=1$.
However, since $a_{k+1}^{l+1}=a_1^{l+1}$ and $a_{k+1}^n=a_1^n$, this contradicts the fact that $a_1^{l+1}\neq a_1^n$.
Thus $a_{k+1}^{l+1}$ and $a_{k+1}^n (n\le l)$ are co-prime.
\qed

\begin{Corollary}
The bilinear dKdV equation \eqref{bilin} with an initial condition
\begin{equation}
a_m^0=1,\ a_0^n=1,\ a_k^1=\left(\prod_{j=1}^k w_j^0\right)^{-1}\ (m,n\ge 0,\ k\ge 1), \label{awcorresp}
\end{equation}
satisfies the following:
\begin{itemize}
\item $a_m^n\in K_w:=K[(w_1^0)^{\pm 1},(w_2^0)^{\pm 1},\cdots, (w_m^0)^{\pm 1}]$,
\item $a_m^n$ is irreducible as an element of $K_w$,
\item $a_m^n$ and $a_{m'}^{n'}$ are co-prime as elements of $K_w$ if $(m,n)\neq (m',n')$.
\end{itemize}
\end{Corollary}
\textbf{Proof}\;\;
The elements of $I_a$ (i.e. $x_1,x_2,\cdots$), and the elements of $I_w$ (i.e. $w_1^0,w_2^0,\cdots$) correspond to each other as \eqref{awcorresp}. Since these relations are given by monomials in $I_a$ and $I_w$, the irreducibility and co-primeness are preserved under the transformation given by \eqref{awcorresp}.
\qed
As a consequence of this corollary, we can obtain the following theorem on the common factors among the terms of the \textit{nonlinear} dKdV equation.
\begin{Theorem} \label{thm1}
Let us consider the nonlinear dKdV equation \eqref{nonlin} with the boundary condition
\[
w_0^n=1\ (n=0,1,2,\cdots).
\]
The terms of the nonlinear dKdV equation are rational functions of the initial values $w_m^0\ (m\ge 1)$ that satisfy the following \textit{co-primeness} condition:

Let us express two terms $w_m^n$ and $w_{m'}^{n'}$ in minimal forms as rational functions, i.e.
\[
w_m^n=\frac{F}{G},\ w_{m'}^{n'}=\frac{H}{K},
\]
where $F,G,H,K$ are polynomials in $w_m^0$ $(m\ge 1)$, and that $F$ and $G$ are co-prime as polynomials, and that $H$ and $K$ are co-prime in the same manner.
Then every pair of polynomials from the four polynomials $F,G,H,K$ does not have common factors other than the monomials in terms of $w_m^0$ $(m\ge 1)$ if
\[
|n-n'|\ge 2\ \ \mbox{or}\ \ |m-m'|\ge 2.
\]
\end{Theorem}
\textbf{Proof}\;\;
The proof of this theorem is immediate from the relation \eqref{relation} and the co-primeness of the terms in the bilinear dKdV equation proved in proposition \ref{prop2}.
\qed
Above theorem \ref{thm1} indicates that the poles and zeros of the nonlinear dKdV equation do not propagate beyond two steps away in each direction. This fact can be interpreted as the confinement of singularities of the equation, and therefore gives the precise description of the singularity confinement criterion.

\section{Concluding remarks and discussions}
In this article we investigated the confinement of singularity of the dKdV equation in terms of the irreducibility and co-primeness of the terms.
Proposition \ref{laurentbilin} established that the terms $a_m^n$ of a solution to the bilinear equation \eqref{bilin} are irreducible Laurent polynomials of initial conditions $I_a$. Co-primeness of distinct terms follows from the proof of proposition \ref{laurentbilin}, hence we conclude that the singularities, the zeros, do not propagate in the evolution of equation \eqref{bilin}.
In proposition \ref{prop2}, we proved the same statements for the bilinear dKdV equation \eqref{bilin} with restricted initial conditions $a_0^n=1,(n\ge 0)$.
Out main result is theorem \ref{thm1}, in which the confinement of singularities of the nonlinear dKdV equation \eqref{nonlin} has been reformulated in terms of co-primeness of the terms.
In this article we have proved the co-primeness of the nonlinear dKdV equation with the boundary condition ($w_0^n=1,(n\ge 0)$), so that we can directly make use of the irreducibility of the bilinear dKdV equation \eqref{bilin}.
We assert that the terms $w_m^n$ and $w_{m'}^{n'}$ of a solution to the nonlinear equation \eqref{nonlin} are co-prime as rational functions of the initial conditions $I_w$ if $|m-m'| \ge 2$ or $|n-n'| \ge 2$, which also means that the singularities, zeros and infinities, are confined.
We needed the boundary condition in order for us to obtain monomial relation between the set of initial conditions $I_a$ of the bilinear dKdV equation \eqref{bilin}, and that $I_w$ of the nonlinear dKdV equation \eqref{nonlin}.
This relation is displayed as \eqref{nonlinbilinrel}.
We conjecture that theorem \ref{thm1} holds for generic boundary conditions. The proof of this conjecture is currently in progress.

As is seen from the proofs, we can expect that similar statements hold for other integrable partial difference equations.
In fact, it is shown that the bilinear form of the Hirota-Miwa equation, which is the master equation of a series of integrable partial difference equations, has the same property as the dKdV equation \cite{Mase,Mase2}.  
Hence we conjecture that the `co-primeness' is an alternative mathematical statement of singularity confinement in nonlinear partial difference equations, and that it gives an integrability criterion for difference equations. 

Some part of the proof of propositions in this paper relies on the fact that the base field $K$ has characteristic $0$.
Therefore, it is quite an interesting task to investigate the singularities of the dKdV equation over the field of positive characteristic such as for finite fields.
It is also necessary to study the relationship of our method to the concept of `almost good reduction' introduced in \cite{Kanki}, which is a formulation of the singularity confinement test in terms of $p$-adic dynamical systems. The almost good reduction criterion is useful for maps of the plane, in particular for the QRT mappings \cite{QRT} and the discrete versions of the Painlev\'{e} equations. However, up to now, the extension of this criterion to higher dimensional maps has not been obtained. This paper is expected to give some insight from the viewpoint of co-primeness of the terms, when we want to construct a generalized reduction property related to the integrability of the mappings.

Finally let us comment on the relationship of our methods with algebraic entropy.
The Laurent property and the integrability detection by the algebraic entropy are closely connected. However, having the Laurent property is not equivalent to having zero algebraic entropy. We expect to obtain refined integrability test in terms of Laurent property,
by adopting co-primeness as an additional criterion.

It is hoped that relations among various useful integrability criteria will be obtained and that a new integrated criterion can be proposed in the future.

\section*{Acknowledgments}
The authors wish to thank Profs. R. Willox, B. Grammaticos and Dr. T. Mase for useful comments.
This work is partially supported by Grant-in-Aid for Scientific Research of Japan Society for the Promotion of Science ($24\cdot 1379$).
%%%%%%%%%%%%%%%%%%%%%%%%%%%%%%%%%%%%%%%%%%%%%%%%%

\end{document}